\documentclass[3p,times,twocolumn]{elsarticle}

\usepackage{ecrc}

\volume{00}

\firstpage{1}

\journalname{Nuclear Physics B Proceedings Supplement}

\runauth{M. Gonz\'alez-Alonso}

\jid{nuphbp}

\jnltitlelogo{Nuclear Physics B Proceedings Supplement}

\usepackage{amssymb}

 \biboptions{compress}

\usepackage[figuresright]{rotating}

\newcommand{\be}{\begin{equation}}
\newcommand{\ee}{\end{equation}}

\newcommand{\ba}{\begin{eqnarray}}
\newcommand{\ea}{\end{eqnarray}}

\newcommand{\bi}{\begin{itemize}}
\newcommand{\ei}{\end{itemize}}

\newcommand{\Br}{\mathrm{Br}}

\begin{document}

\begin{frontmatter}



\dochead{}

\title{TAU2014\ Opening Talk}


\author{Martin Gonz\'alez-Alonso}

\address{IPNL, CNRS/IN2P3, 4 rue E. Fermi, 69622 Villeurbanne cedex, France; Universit\'e Lyon 1, Villeurbanne\\
Universit\'e de Lyon, F-69622, Lyon, France}

\begin{abstract}
Different aspects of Tau physics are discussed, both as a New Physics probe and as a tool to study Standard Model physics. 
The goal of this Tau2014 opening talk is to introduce (some of) the many research directions currently being pursued with tau leptons, which will be discussed in great technical detail in the following talks.
\end{abstract}

%
%

\end{frontmatter}

\section{Introduction}

Almost four decades after the discovery of the $\tau$ lepton at the SPEAR $e^+e^-$ storage ring\footnote{Martin Perl, who won the Nobel prize for this discovery, sadly passed away few days after this Tau2014 conference.}~\cite{Perl:1975bf} we have accumulated a large amount of precise experimental information about its properties and its many decay channels \cite{Agashe:2014kda,Amhis:2012bh}. On the other hand, theoretical efforts have followed suit, improving significantly our understanding of the $\tau$ dynamics. For a detailed recent review of these developments see Ref.~\cite{Pich:2013lsa}.

The physics of the tau lepton are exceptionally broad, both experimentally and theoretically, as nicely shown by the Tau2014 agenda:
\begin{itemize}
\item We will start the conference learning about the latest measurements of the basic tau properties, such as mass, lifetime, etc. 
\item Then we will look at the tau lepton as a low energy QCD laboratory thanks to its hadronic decays, where we can distinguish three main (overlapping) efforts: (i) exclusive decays; (ii) inclusive decays; (iii) and the relation with the muon magnetic dipole moment. 
\item The third day of the conference will focus on the violation of the lepton flavor (LFV), both in the charged sector, where it represents a superb New Physics (NP) probe, and in the neutral sector, where we find the rich field of neutrino physics. 
\item On Thursday we will move to higher energy scales, making the tau a light particle generated through the decay of heavy quarks or the proton collisions at the LHC. 
\item At last we will end the conference discussing future facilities that will lead the field in the coming years. 
\end{itemize}

Such a broad spectrum of topics entails a very heterogeneous list of participants, some of whom are probably not very familiar with certain topics. One of the goals of this opening talk is to give a not too technical introduction to (some aspects of) the field, with the hope that it will make easier to follow the subsequent discussion of recent developments.

The outline of this talk is the following: Section \ref{sec:leptonic} discusses leptonic $\tau$ decays as New Physics probes, Section \ref{sec:hadronic} introduces the theoretical framework used to describe inclusive hadronic tau decays and Section \ref{sec:LFV} presents LFV searches with tau leptons and their theoretical implications. Such an outline is strongly influenced by my own work in the field, and needless to say, it is by no means comprehensive. Nonetheless the topics not covered in this introductory talk (exclusive hadronic decays, neutrino physics, colliders, ...), will be thoroughly reviewed in the coming days of the conference.

\section{Leptonic Tau decays}
\label{sec:leptonic}
\subsection{Lepton universality tests}
The Standard Model (SM) treats all lepton flavors equally up to usually small effects coming from their different masses. Such prediction can be experimentally tested by the so-called lepton universality tests, through the comparison of processes that only differ in the flavor of the leptons involved. In order to compare the sensitivity of different tests it is convenient to work with a flavor dependent $W\ell\nu_\ell$ coupling $g_\ell$, and to derive experimentally the value of the ratios $g_\ell/g_{\ell'}$.\footnote{It is worth stressing that the violations of lepton universality can have a different origin than the simple modification of the $W\ell\nu_\ell$ vertex.}

The most precise determination of $g_e/g_\mu$ comes from tau decays and pion decays, both at the $0.15\%$ level and in good agreement with universality. Slightly weaker limits are obtained with kaon and Z decays \cite{Amhis:2012bh,Pich:2013lsa}.

Taking into account the recent measurements of the muon and tau lifetimes performed by the MuLan and Belle collaborations~\cite{Webber:2010zf,Belle}, similar precisions are achieved for $g_\ell/g_\tau$ ($\ell=e,\mu$) using tau leptonic decays. Slightly weaker limits are derived from semileptonic tau decays and leptonic Z decays \cite{Amhis:2012bh,Pich:2013lsa}. And once again, all these per-mil level measurements are in good agreement with universality.

There are however two measurements that are in a small tension with universality: (i) the old LEP2 anomaly in W decays; (ii) and the more recent anomaly in $\bar{B}\to D^{(*)}\ell\nu_\ell$. Let us take a closer look at these two measurements.
\subsubsection{$W\to\ell\nu$}
Measurements of the leptonic $W$ decays performed at LEP2 can be used to calculate the ratio
\cite{Agashe:2014kda,Alcaraz:2006mx} 
\ba
\frac{2 \, \mbox{BR} \left(W \rightarrow \tau \, \overline{\nu}_{\tau} \right)}{\mbox{BR}  \left( W \rightarrow e \, \overline{\nu}_{e} \right) + \mbox{BR} \left(W \rightarrow \mu \, \overline{\nu}_{\mu} \right)}
= 1.055 (23),
\nonumber
\ea
which happens to be $2.4$ standard deviations\footnote{The anomaly becomes $2.8\sigma$ using only LEP2 data \cite{Alcaraz:2006mx}.} (all correlations included) away from the SM prediction $R^W_{\tau\ell}|_{\mbox{\tiny SM}}= 0.999$ \cite{Kniehl:2000rb}, which uncertainty is negligible compared with the experimental error. 

From the above discussion it is clear that a simple modification of the W vertex at the per-cent level (required to accommodate this anomaly) is not acceptable since it would spoil the per-mil level agreement of lepton universality in tau, meson and Z decays. This simple observation is the main reason why this anomaly has not received a lot of attention by model-builders (see Refs.~\cite{Li:2005dc,Park:2006gk,Dermisek:2008dq} for a few exceptions).

This argument is in principle invalidated by the introduction of additional lepton flavor violating interactions that can interfere with the modified W vertex in tau and Z decays. The room for such a cancellation was studied in Ref.~\cite{Filipuzzi:2012mg} assuming that NP effects at the EW scale can be parametrized by series of higher-dimensional effective operators~\cite{Leung:1984ni,Buchmuller:1985jz}
\ba
{\cal L}_{eff} = {\cal L}_{SM} + \frac{\alpha_i}{\Lambda^2}{\cal O}_6 + \ldots ~,
\label{eq:SMEFT}
\ea
where not only the SM gauge symmetries, but also the flavor symmetry $[U(2)\times U(1)]^5$ was assumed. Somehow surprisingly it was found that not even in such a general scenario (with 17 new parameters) it is possible to accommodate the LEP2 anomaly in the leptonic W decays. 

It is worth stressing that this Effective Field Theory (EFT) analysis assumes that the LEP anomaly comes from a non-standard W vertex and it does not explore possible contributions of different origin affecting the $e^+e^-\to \tau \nu_\tau\ell \nu_\ell,~ \tau \nu_\tau qq'$ processes, which would contaminate the measurement of the $\rm{W\to\tau\nu_\tau}$ vertex.\footnote{I thank Achim Stahl for useful comments concerning this point.}

An orthogonal possibility is represented by New Physics models containing light (compared with the weak scale) new particles, such as in Ref.~\cite{Park:2006gk,Dermisek:2008dq} where charged Higgses with a mass similar to the W mass are present, generating an apparent violation of lepton flavor universality in the W vertex.

Needless to say, the effect might just be a statistic or systematic effect in the measurement. In this respect it is worth noting that although all four LEP collaborations found an excess of tau events, the anomaly is significantly dominated by the measurement of the L3 Coll.
\subsubsection{$\bar{B}\to D^{(*)}\ell\nu_\ell$}
In 2012 BABAR reported the following results concerning $b\to c\,\ell^-\bar\nu_\ell$ transitions \cite{Lees:2012xj}
\ba
R(D)&\!\! \equiv &\!\!
\frac{\Br(\bar B\to D\,\tau^-\bar\nu_\tau)}{\Br(\bar B\to D\,\ell^-\bar\nu_\ell)}
\; = \; 0.440(72)\, ,\\
[5pt]
R(D^*)&\!\! \equiv &\!\!
\frac{\Br(\bar B\to D^*\tau^-\bar\nu_\tau)}{\Br(\bar B\to D^*\ell^-\bar\nu_\ell)}
\; = \; 0.332(30)\, ,
\ea
where $\ell=e,\mu$. These results disagree with the SM prediction, $R(D)=0.296(17)$ and $R(D^*)=0.252(4)$ ~\cite{Lees:2012xj,Kamenik:2008tj,Fajfer:2012jt}, by $2.0\sigma$ and $2.7\sigma$ respectively ($3.4\sigma$ taken together). Previous Belle measurements~\cite{Bozek:2010xy} are less precise and agree both with the SM predictions and the BABAR results.

The violation of lepton flavor universality suggested by these results is enormous, namely ${\cal O}(30\%)$. However, in contrast to the situation with the LEP2 anomaly, these tensions can be interpreted as the consequences of an effective four-fermion interaction, which avoids conflict with the per-mil level agreement found in other lepton universality tests.  

This has triggered an intense theoretical activity to study how to generate such four-fermion interaction in different extensions of the SM, without re-introducing the problem in other heavy quark (semi)leptonic decays. These efforts will be covered with some detail in A. Celis's talk~\cite{Celis}.

On the other hand the SM prediction for $R(D)$ has been revisited recently using new approaches~\cite{Bailey:2012jg,Becirevic:2012jf}, finding somewhat larger values that decrease the tension with the BABAR measurement below $2\sigma$. Similar studies of $R(D^*)$ are expected in the near future~\cite{Bailey:2012jg}.

On the experimental end it will be interesting to learn later in this conference about the status of analyses of $R(D)$ and $R(D^*)$ using the full Belle data sample~\cite{Kuhr}. A confirmation (or increase) of the tension with the SM prediction would motivate the measurement of the corresponding differential distributions, which would provide an excellent handle to understand the origin of the discrepancy~\cite{Celis}.

\subsection{Michel parameters}

Beyond the measurement of branching ratios and lifetime, which offer the tests of  lepton flavor universality discussed above, there are additional observables that can be studied in leptonic tau decays, such as angular asymmetries or the energy spectrum of the final lepton. These observables can be used to study the Lorentz structure of the interaction, which in the SM is purely $\rm{V-A}$. Deviations from the SM result generated by new particles which are much heavier than the tau lepton can be incorporated through the following low-energy effective Lagrangian
\ba
\!\!\!\!\!\!\!\!\!{\cal L}_{eff} = - 4 \frac{G_{\tau\ell}}{\sqrt{2}}
\sum_{\epsilon,\omega = R,L}^{n = S,V,T}
\!\!g^n_{\epsilon \omega}
\left[ \overline{\ell_\epsilon}
\Gamma^n {(\nu_\ell)}_\sigma \right]\,
\left[ \overline{({\nu_\tau})_\lambda} \Gamma_n
	\tau_\omega \right],
\label{eq:tauEFT}
\ea
which contains nineteen independent real parameters for each leptonic channel, of which only one specific combination is probed through 
lepton universality tests. The study of the angular and energy distribution of polarized tau leptons, $d^2\Gamma / dx\,d\!\cos\theta$, provides access to four additional combinations for each channel, usually denoted by $\rho, \eta, \xi, \delta$ and known as Michel parameters~\cite{Agashe:2014kda}.

Their current experimental values, which come from LEP and CLEO data~\cite{Agashe:2014kda}, are known with a $1\!-\!5\%$ precision and are in agreement with the SM values $\rho=\delta=3/4, \eta=0$ and $\xi=1$. An order of magnitude improvement is expected from the analysis of Belle data~\cite{Epifanov}.

These results impose non-trivial constraints on the Wilson coefficients $g_{\epsilon\omega}^n$, which in a specific NP theory are just functions of the new couplings and masses. Unfortunately the sensitivity to the NP contributions of all Michel parameters is quadratic, except in the case of $\eta$, which effect in the distribution is very suppressed by the factor $x_0=m_\ell/E_{max}$ except in a small region at the beginning of the spectrum. 

Matching the low-energy EFT at the tau scale given by Eq.~(\ref{eq:tauEFT}) with the EFT at the EW scale given by Eq.~(\ref{eq:SMEFT}) opens the possibility of studying the complementarity of these searches with collider studies, and with low-energy neutral-current processes (connected through the $SU(2)$ gauge symmetry) \cite{Cirigliano:2009wk}. Interestingly enough, it is found that only one weak-scale effective operator contributes to the $\eta$ parameter \cite{Cirigliano:2009wk,MGAthesis}, namely
\ba
\eta = \frac{1}{2}\rm{Re}\left( g_{RR}^S g_{LL}^{V*} \right) \approx \frac{1}{2}\rm{Re} \,g_{RR}^S = \rm{Re} \, \left[\alpha_{\ell e} \right]_{3ii3} \frac{v^2}{\Lambda^2}~.
\nonumber
\ea
where\footnote{Here $\ell~(e)$  is the left(right)-handed $SU(2)$ lepton doublet (singlet) with three components in flavor space. Thus, both the operator ${\cal O}_{\ell e}$ and its Wilson coefficient $\alpha_{\ell e}$ have four flavor indexes.} 
 ${\cal O}_{\ell e}\equiv \bar{\ell}\gamma_\mu \ell \cdot \bar{e}\gamma^\mu e$ and $v\approx 174$ GeV is the vacuum expectation value of the Higgs field. In the electronic channel ($i=1$), the corresponding effective operator affects also the LEP2 process $e^+e^-\to\tau^+\tau^-$, which presumably will set strong bounds on the Wilson coefficient\footnote{Usual EFT analyses of LEP observables assume $U(3)^5$ or $U(2)^5$ flavor symmetry, and therefore do not have this operator.}$^,$\footnote{Notice that the same argument applies to the $\mu$-decay $\eta$ parameter.} Even so, it is remarkable that $\eta$ probes only one operator, whereas LEP2 probes combinations of operators, offering a possible complementarity. On the other hand, the muonic tau decay ($i=2$) probes an operator which is not probed by LEP, providing a window into an unexplored direction in the EFT parameter space.

The discovery of the Higgs, along with the absence of new phenomena at the LHC, has increased the interest in this kind of EFT analysis. In that context, it is interesting to identify the operators probed by this kind of experiments and the associated current/future limits.

Finally let us mention that the measurement of the polarization of the final lepton would open interesting possibilities. On one hand it would make possible to construct CPV triple products, and access the imaginary part of $g_{RR}^S$. On the other hand, the above-mentioned mass suppression affecting the real part of $g_{RR}^S$ in the $\eta$ term would not be present, offering possible better bounds. 

\section{Inclusive hadronic Tau decays}
\label{sec:hadronic}
Inclusive hadronic tau decays represent an interesting arena to study low-energy QCD and extract fundamental parameters of the SM with great accuracy. The theoretical framework needed to do so can however appear opaque to the non-practitioners. In order to ease this situation, this Section will introduce the main conceptual elements of this framework and identify the different points subject to current scrutiny. The goal is to set the stage for the subsequent presentations and discussion that we will have in the corresponding session of this Tau2014 conference, where the different development of the field will be explained, this time with great technical detail.
\subsection{A brief introduction}
The study of inclusive QCD observables requires by definition the injection of a quark current in the QCD vacuum, with enough energy to create hadrons. $e^+e^-$ collisions can inject a vector current (the Z-component is swamped by the photon-component at low-energies), whereas the hadronic decays of the tau lepton offer a clean way of injecting both vector and axial-vector current. This makes possible e.g. the study of the spontaneous breaking of the chiral symmetry, a purely non-perturbative QCD effect.

As we will see, the fact that $m_\tau>\Lambda_{QCD}$ turns to be crucial, because it will allow the use of analytical methods to describe inclusive observables. On the other hand, nature could have been even kinder, giving us a heavier tau lepton, since a larger separation of scales would have made theorists' work much easier.

On the experimental end it is important to stress that LEP collaborations were able to measure not only the branching ratios, but also the energy distributions of the hadronic systems (i.e. spectral functions). As we will see below, the experimental knowledge of these functions makes possible to construct tailor-made sum rules depending on the goal, which triggered an enormous QCD activity during these last two decades.

First of all, let us start by noting that the central theoretical object describing an inclusive observable is the appropriate two-point correlation function:
\ba
\nonumber
\Pi_J^{\mu\nu} (q)
\!\!\!&\equiv&\!\!\! i\int d^4 x\, e^{iqx} \langle 0 | T\left[ J^\mu (x) J^\nu (0)^\dagger \right] | 0 \rangle \\
\nonumber
\!\!\!&=&\!\!\! (-g^{\mu\nu}q^2 +q^\mu q^\nu) \Pi_J^{(1)} (q^2)    +    q^\mu q^\nu \Pi_J^{(0)} (q^2) ~,
\ea
where $J_\mu$ is the quark current injected in the QCD vacuum. It is well-known that the corresponding spectral functions are given by the discontinuity of the correlators $\Pi_J^{(0,1)} (q^2)$ in the positive real axis, a property based only on translation invariance, unitarity and microcausality.

%
\begin{figure}[t!]
\centering
\includegraphics[width=0.4\textwidth]{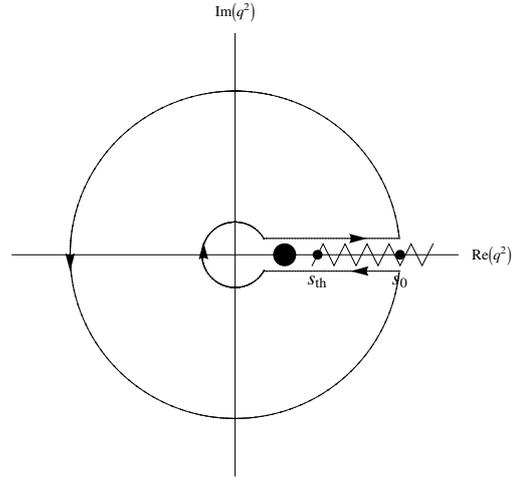}
\caption{Circuit of integration in the $q^2$-complex plane. The radius of the inner circle is $\epsilon$, the radius of the outer one is $s_0$ and the separation between the horizontal parts of the circuit is proportional to $\epsilon$. We indicate the possible pole of the correlator at $q^2\leq s_{th}$ and the cut starting at $q^2=s_{th}$. }
\label{fig:circuit}
\end{figure}
%
The so-called QCD sum rules exploit the analytic properties of a given correlator $\Pi(q^2)$ to relate data and theory \cite{Shifman:1978bx}. The application of this framework to the description of inclusive hadronic tau decays started in the late 80s (see e.g. Ref.~\cite{Peccei:1986hw,BNP92,Le Diberder:1992fr}), and it has been refined and applied to the extraction of different quantities ever since. The main idea is to integrate the correlator times a weight function\footnote{The weight function is assumed to be analytic except for a possible pole at the origin.} $w(z)$, over the contour C shown in Fig.~\ref{fig:circuit}. By Cauchy's theorem we have
\ba
\oint_C \Pi(z)~w (z) = 0~.
\nonumber
\ea
The conceptual content of this \emph{exact} relation becomes clear when the contour is decomposed in three parts with very different physical content:
(i) the part around the positive real axis, which only non-zero contribution comes from the discontinuity of the correlator, i.e. the observable hadronic spectral function; (ii) the small circle around the origin, where Chiral Perturbation Theory (ChPT) applies; and (iii) the outer contour, where the integral of the correlator appears. 

At this point, and in order to make this relation useful phenomenologically, we are forced to make a key \emph{approximation}: we will use the Operator Product Expansion (OPE) to calculate the correlator in the contour. This makes the resulting expression a non-trivial relation between quarks (theory, QCD) and hadrons (experiment), a sort of duality. For this reason the failure of this approximation is called quark-hadron duality violation (DV). It is important to emphasize that DV is in general non-zero for finite values of the Cauchy's radius\footnote{To make things more explicitly, one can take $s_0 \approx m_\rho$ and see how the resulting sum rule will have a enormous DV error.}, but its precise numerical value is unknown in any given case, beyond the naive expectation that it will be small if $s_0\gg \Lambda_{QCD}$. 
Such a simple observation is pertinent, since the literature is some times confusing about the existence of DV. Such an existence must be always understood as numerical relevance for a specific case.

All in all  we have derived the expression of a general QCD sum rule
\ba
\nonumber
\oint_{|z|=s_0} \!\!\!\!\!\Pi_{OPE}(z)~w (z)\,dz - \oint_{|z|=\epsilon}\!\!\!\!\Pi_{\chi PT}(z)~w (z)\,dz \\
+ \int_{s_0}^\infty\!\!\! \rho (s)\,w(s)\,ds = - DV\left[\Pi\,w, s_0\right] ~.
\nonumber
\ea
All four terms in this expression are subject to intense current investigations and we will learn about recent developments in the coming days. Let us say some brief comments about them.

\textbf{The OPE term} contains a perturbative 
 and a non-perturbative piece .
 The perturbative contribution, being the dominant, needs to be calculated with great accuracy. Impressively enough, results at ${\cal O}(\alpha_s^4)$ are available \cite{Baikov:2008jh}. The main error comes from the treatment of higher order corrections, through the use of different renormalization-group improvement prescriptions in the integration~\cite{Le Diberder:1992te,Beneke:2008ad,Menke:2009vg,Caprini:2009vf,DescotesGenon:2010cr,Cvetic:2010ut,Beneke:2012vb,Abbas:2013usa}. 
On the other hand, the non-perturbative contribution is encoded in the OPE by the so-called vacuum condensates, which are suppressed by powers of $\Lambda_{QCD}/s_0$. Both their phenomenological extraction and the numerical impact of neglected higher-dimensional condensates in any given sum rule have been subject of recent efforts, see e.g. Refs.~\cite{Maltman:2008na,Maltman:2008nf,GPP2010,Bodenstein:2011hm,Boito:2012nt,Davier:2013sfa}.

\textbf{The ChPT term} is present if the weight function has a pole at the origin. The challenges here concern the calculation of the appropriate correlator at a given order in ChPT (see e.g. Ref.~\cite{Amoros:1999dp,GonzalezAlonso:2008rf}), and their phenomenological implementation, given our limited knowledge of some of the NNLO low-energy constants involved~\cite{Boito:2012nt,GonzalezAlonso:2008rf,Boyle:2014pja,Golterman:2014nua}.

The determination of the different \textbf{spectral functions} has been subject of an intense experimental activity during the past two decades~\cite{Ackerstaff:1998yj,Davier:2005xq}. Concerning recent developments, ALEPH data was updated correcting the unfolding procedure~\cite{Davier:2013sfa,Boito:2010fb} and B-factories have provided data for specific channels \cite{Hayashii}. However, a determination of inclusive spectral functions from the B-factories has not been done, despite having much more statistics than LEP collaborations. 

At last we have \textbf{the DV component}, which has proven to be a very complicated theoretical problem.\footnote{See Refs.~\cite{Shifman:1998rb,Shifman:2000jv} by M.~Shifman for very enlightening discussions concerning DV.} Nonetheless, on the phenomenological side DV has received significant attention during the last few years, at least in the context of hadronic tau decays. Different works have tried to quantify its numerical importance for specific sum rules, through the use of specific models or parametrizations, see e.g. Refs.~\cite{GPP2010,Cata:2008ye,Peris}.

After this general discussion it might be useful to close this subsection with some specific examples:
	\begin{itemize}
	\item The V+A non-strange correlator and its associated spectral function can be used to determine the strong coupling constant with great accuracy~\cite{BNP92,Davier:2013sfa,Peris,Pich}. The comparison with $\alpha_s$ extractions at higher energies provides a beautiful test of asymptotic freedom.
	\item The comparison of V+A non-strange and strange channels can be used to determine the CKM element $V_{us}$~\cite{Vus,Lusiani} with an accuracy close to the best current extraction from kaon decays.\footnote{Interestingly enough, current $V_{us}$ determination from inclusive tau decay are in $\sim 3\sigma$ tension with unitarity~\cite{Lusiani}.} This provides, together with the $V_{ud}$ determination, a powerful NP test, sensitive to 10 TeV effective scales and complementary to collider searches~\cite{Cirigliano:2009wk}.
	
	\item The comparison of non-strange vector and axial-vector channels can be used to determine non-perturbative quantities, such as the ChPT low-energy constants $L_{10}$ and $C_{87}$ or the corresponding $D=6,8,...$ vacuum condensates~\cite{Boito:2012nt,GonzalezAlonso:2008rf,Boyle:2014pja,Golterman:2014nua}.
	\end{itemize}
\subsection{Connection with $(g-2)_\mu$}
Let us briefly discuss now the connection between hadronic tau decays and the anomalous magnetic moment of the muon $a_\mu = (g-2)_\mu/2$. A more detailed discussion will be given in the talk by P.~Marquard~\cite{Marquard}.

Currently the main source of error in the SM theoretical prediction of $a_\mu$ comes from the hadronic vacuum polarization (HVP) contribution, followed by the so-called light-by-light contributions (LBL) \cite{Masjuan}. The leading HVP can be expressed as the following integral (see e.g. Ref.~\cite{Jegerlehner:2009ry})
\ba
a_\mu^{HVP} = \alpha^2 \, m_\mu^2 \int_{0}^\infty ds~ \rm{Im}\,\Pi_{J_{em}}(s)~w(s)~.
\ea
The kernel $w(s)$ has an approximate $1/s^2$ behavior that gives very high weight to the low-energy region, around the $\rho$ resonance. Although calculations of the correlator in that region from first principle are not available, it is possible to determine its value phenomenologically, either from $e^+e^-\to$ hadrons (via scan or ISR method), or from tau decay (through an isospin rotation). Current determinations using these two approaches are not fully consistent, and to make things more complicated there are also some discrepancies between different $e^+e^-$ experiments.\footnote{A sum rule with the appropriate weight can reduce the data dependence, but introducing certain OPE vacuum condensates~\cite{Bodenstein:2013flq}.}
 In the next days there will be several talks describing the recent developments in all these fronts.

The precise determination of this contribution has become a very important issue because there is a 2-3 sigma tension between the current experimental value for $a_\mu$ and the SM theoretical prediction~\cite{Blum:2013xva}, which could represent a hint of New Physics.

It should be also noticed that lattice efforts are ongoing to calculate the HVP contribution~\cite{Maltman}.

To set the stage for the subsequent discussion in the coming days, let me finish showing the status of these calculations after the last Tau conference~\cite{Pich:2013kg}
\ba
a_\mu \times 10^{10}
&=& 11\,658\,471.9\,(0.0)~~~~(QED)~,\nonumber\\
&& ~~~~~~~~~+15.4\,(0.2)~~~~(EW)~,\nonumber\\
%
%
&&
\left\{
\begin{array}{ll}
~~~+692.4\,(4.1)&(hvp~e^+e^-)~,\\
~~~+701.5\,(4.7)&(hvp~\tau)~,
\end{array}
\right.
\nonumber\\
&& ~~~~~~~~~~~~-9.8\,(0.1)~~~(hvp~NLO)~,\nonumber\\
&& ~~~~~~~~~+10.5\,(2.6)~~~~(LBL)~,\nonumber\\
&=&
\left\{
\begin{array}{ll}
11\,659\,180.4(4.9)&(e^+e^-)~,\\
11\,659\,189.5(5.4)&(\tau)~,
\end{array}
\right.
\nonumber
\ea
to be compared with the experimental value $\rm{a_\mu=11\,659\,208.9(6.3)}$~\cite{Bennett:2006fi}.

\subsection{Higgs as a QCD laboratory?}
Similarly to the tau lepton, which is able to inject a quark current into the QCD vacuum with energies between zero and the tau mass, we could use the recently discovered Higgs boson, through its decays $h\to ZZ^*\to\ell^+\ell^-q\bar{q}$ and $h\to WW^*\to\ell^-\nu u\bar{d}$, with the lepton pair coming from the on-shell Z/W.\footnote{The relevant QCD correlators affect also the $h\to 4\ell$ channel \cite{Gonzalez-Alonso:2014rla}. Although the effect occurs in this case through one-loop diagrams, it is much cleaner experimentally.} 

The theoretical description goes along the same lines as the hadronic tau decays and so it benefits from the progress achieved in that field. It would also shed some light on issues such as DV since the Cauchy's radius can now be much larger than the tau mass, namely $s^{max}_0=(M_h-M_z)^2\approx (35 \mbox{GeV})^2$ if the $Z$ decaying to leptons is exactly on-shell.

The challenge is however on the experimental side. In order to be able to access the relevant low-energy region of the spectrum with a small enough bin size, very large statistics are necessary, along with a careful control of the background. Even at a high-luminosity LHC these requirements seem too hard (current $h\to4\ell$ data contains only a few dozens of events!), but perhaps a future $e^+e^-$ facility could offer some useful results.

\section{Lepton flavor violation}
\label{sec:LFV}
\subsection{LFV tau decays}
Lepton flavor is an accidental symmetry in the SM if neutrino masses are neglected, and the LFV effects due to the finite neutrino masses are completely negligible for practical purposes~\cite{Lee:1977tib}, which makes these searches the ideal smoking gun for new physics.

Once we go beyond the Standard Model (BSM) this accidental symmetry is not expected in general, and indeed it does not happen in many NP models, which easily generate LFV signals close to current experimental limits.

The enormous experimental efforts along with the absence of a SM contribution make current experiments sensitive to LFV-interactions generated by particles much heavier than the scales directly accessible at colliders. 
This represents an important source of information about the BSM physics, which NP models need to explain or at least accommodate.

LFV in the muon/electron sector is much more constrained experimentally than in the tau sector, but the comparison between them is only possible in specific NP theories scenarios. In this sense, such comparison is a model-diagnostic tool, which tell us about the flavor structure of the new theory. Similarly, the different channels (e.g. $\tau\to\mu\gamma$ vs. $\tau\to\ell\ell\ell$) have also a strong diagnostic power, since the underlying LFV interactions affecting each observable are in general different. See review talks by Y.~Uchida~\cite{Uchida} and E.~Passemar~\cite{Passemar} for more details.

The B-factories have improved by more than an order of magnitude previous bounds by CLEO in a long list of LFV tau decays~\cite{Amhis:2012bh,Hayasaka} and a similar improvement is expected from Belle-II with 50 fb$^{-1}$. 
It is worth noticing that in specific channels, such as $\tau\to3\mu$, the bounds obtained by the LHCb collaboration are becoming competitive with those from the B-factories~\cite{Chrzaszcz}.  

\subsection{LFV Higgs decays}
An interesting front has been opened after the discovery of the Higgs boson, since its decays could be LFV once we abandon the SM framework, and since once again they appear quite naturally in BSM theories.

Indirect bounds on LFV Higgs couplings can be derived, since they induce muon or tau LFV decays. Such limits are indeed very strong for $h\to\mu e$, making it unobservable at the LHC, but the limits for the channels $h\to\tau \ell~(\ell=\mu,e)$ are much weaker, with branching ratios at the 10\% level still allowed \cite{Kanemura:2005hr,Davidson:2010xv,Blankenburg:2012ex, Harnik:2012pb}. This gives the LHC clear discovery opportunities in these channels.

A couple of years ago a 13\% limit on $B( h\to\tau \ell )$ at 95\%C.L. was obtained from a theorist reinterpretation of the $h\to\tau\tau$ data~\cite{Harnik:2012pb}, and it was pointed out soon afterwards that a dedicated analysis using 20fb$^{-1}$ of $\sqrt{s}= 8$ TeV data could reach sensitivities below 1\%~\cite{Davidson:2012ds}.

Indeed, such precision has recently (only 2 months ago!) been achieved by CMS in the $\mu\tau$ channel, in a work that represents the first dedicated analysis by an LHC collaboration~\cite{CMS:2014hha}. Interestingly enough, a slight excess of events with a significance of $2.5\sigma$ is observed. Interpreted as a limit one obtains
\ba
B( h\to\tau \mu ) < 1.57\%~~(95\%C.L.)~.
\ea
It will be interesting to keep an eye on the evolution on this result, since this is just the beginning of a long term program.

Finally it should be mentioned as well that competitive bounds on LFV $Z$ decays can also be obtained at the LHC~\cite{Davidson:2012wn,Aad:2014bca}.

\section{Summary}
Tau physics is a very wide and rich field, where one can learn about low-energy QCD and the dynamic of its bound states, obtain stringent bounds on New Physics, study the neutrino sector or use it a superb tool in collider searches.

Currently there are a few intriguing anomalies with a significance of $2-3\sigma$, like in some lepton universality tests, the $V_{us}$ determination or the recently performed measurement of $B(h\to \tau\mu)$. They deserve further experimental and theoretical scrutiny to assess their robustness as New Physics hints and their interplay with other measurements in specific New Physics frameworks.

The experimental activity is very rich, with many results recently obtained both by the LHC collaborations and by low-energy experiments like BABAR, Belle or BESIII. On the theoretical side, the activity is also intense, both within the SM, where the tau lepton keeps representing a unique QCD laboratory, and beyond the SM to understand the implications of the many measurements being performed.  

Without further ado it is my pleasure to give the floor the rest of speakers. 


\section*{Acknowledgements}
I would like to thank the local organizers for this interesting conference, and S.~Davidson and A.~Pich for helpful discussions and comments on the manuscript. The project was partially funded by the Lyon Institute of Origins, grant ANR-10-LABX-66.







\end{document}